\documentstyle[preprint,eqsecnum,aps,epsfig,tighten]{revtex}

\begin{document}

\title{Two-pion exchange and strong form-factors in covariant field theories}
\author{G.\ Ramalho$^1$, A.\ Arriaga$^1$ and M.\ T.\ Pe\~na$^{1,2}$ \\
\vskip 2mm
{\small $^1$Centro de F{\'\i}sica Nuclear da Universidade de Lisboa, 
1699 Lisboa Codex, Portugal \\
        $^2$Departamento de F{\'\i}sica, Instituto Superior T\'ecnico, 
1096 Lisboa Codex, Portugal}\\[2mm]
}
\date{\today}
\maketitle

\begin{abstract}
In this work improvements to the application of the Gross equation 
to nuclear systems are tested. In particular we evaluate 
the two pion exchange diagrams, including the crossed-box
diagram, using models developed within the 
spectator-on-mass-shell covariant formalism. 
We found that the form factors used in these models induce 
spurious contributions that 
violate the unitary cut requirement. We tested then
some alternative form-factors in order to preserve the unitarity condition. 
With this new choice, the difference between the exact and the 
spectator-on-mass-shell amplitudes is of the order of the one boson 
scalar exchange, supporting the idea that this difference may be parameterized 
by this type of terms.
\end{abstract}

\section{Introduction}
\label{sec1}

Electron scattering experiments off light nuclei at high
momentum transfer (few GeV/c) are expected to be performed at facilities
such as TJNAF. 
These experiments will provide new information on the 
intermediate and short range behavior of the nuclear interaction, 
establishing an important challenge to 
theoretical models. 
In fact, at these regimes, theoretical descriptions of the 
adequate hadronic degrees of freedom -- nucleons, mesons and resonances 
-- have to be based on both relativistic
kinematics and dynamics. 

Over the past years different relativistic formulations have been 
investigated for application to light nuclei, which can
be classified in two major categories:
Relativistic Hamiltonian Dynamics and Relativistic 
Field Theories. The first one, not considered in the present 
paper, assumes basically one of two alternative forms,
the light-front form \cite{keister} and
the instant form. The latter is developed within a Schr{\"{o}}dinger equation 
framework, uses variational Monte Carlo techniques and has been 
applied recently to the 3- and 4-nucleon systems \cite{Urbana98}.  
In this formalism relativistic covariance is achieved through the 
Poincar\'e Group algebra, and the resulting Hamiltonian consists of 
a relativistic kinematic energy operator 
and 2- and 3-body interactions. 
The 2-body interaction, containing asymptotic relativistic 
corrections, is parameterized in order to reproduce the NN data 
($\chi^2 \simeq 1$ for $\sim 30$ parameters) below 350 MeV;
it also comprises boost corrections 
up to the order of $({\bf P}/2M)^2$, where ${\bf P}$ is the total 
3-momentum of the interacting pair and $M$ is the nucleon mass. 
Finally, the 3-body interaction is fitted to the triton binding energy.
The generalization to heavier systems is in principle possible. 

The second category methods are based on meson exchange 
diagrams. The non-perturbative character of the nuclear interaction 
definitely requires infinite sums of these amplitudes. 
As well known, infinite series can effectively be summed by means of 
integral equations, which assigns to the Bethe-Salpeter equation \cite{BetheS} 
a predominant role in the relativistic description of the 
NN problem. 

The Bethe-Salpeter equation is a manifestly covariant 4-dimensional 
integral equation. 
In principle, its interaction kernel should include all 
irreducible diagrams derived from a considered Lagrangian, which is 
clearly out of reach of the present calculations. 
This limitation determines inevitably the truncation of the kernel 
expansion and only a restrict set of appropriate meson exchange 
diagrams is kept.   
 
Even with a truncated kernel the Bethe-Salpeter equation is, beyond ladder 
approximation, very hard to solve. This difficulty lead to the 
development of covariant dimensional reductions, imposing restrictions 
on the energy variable, from which 3-dimensional 
integral equations, known as Quasi-Potential (QP) equations,
are obtained. 

Among QP equations, the Gross equation \cite{Gross69} considers only
ladder and crossed-ladder diagrams, and 
is derived from the assumption that important cancellations
between the two categories of
diagrams occur. These cancellations mean that their sum is dominated 
by the contribution, to the ladder diagrams, of the mass pole of the
heavier particle or, 
for equal interacting masses, by the combination of the 
corresponding mass poles. 
In the first case, the Gross prescription consists on replacing 
the sum of the ladder and crossed-ladder diagrams by the ladder
diagrams with the   
massive particle on its mass shell. 
In the second case, a symmetrized combination of the ladder diagrams 
with one of the interacting particles on its mass shell 
is required. The amplitudes obtained contain ($\sim 13$) parameters 
(meson masses and coupling constants) 
fitted to the NN data 
($\chi^2 \simeq 2-3$) below the pion production threshold.

For scalar interacting particles 
in the static limit (the limit 
when the mass of one of the interacting particles goes to infinity), 
the above cancellation is proved to be exact order by order 
\cite{Gross69,GrossBook}, meaning that in this case the Gross prescription 
is exact.  When one of the masses, although finite, remains 
much larger than the other, the prescription is a very good approximation 
as shown by Maung and Gross \cite{Gross90}. 
This result motivates the application of the Gross or spectator 
equation to the proton-nucleus scattering, such as 
$p-^{40}$Ca scattering \cite{Maung89}. 
For equal masses it may be shown 
that the approximation is still reasonably good 
\cite{Ramalho98}.   

In nuclear applications, however, the spinor structure of the nucleon 
is essential, and the vertex interaction includes a Dirac matrix structure. 
Moreover, for pion exchanges with pseudovector coupling (PV),
the coupling supported  by chiral symmetry arguments,
the integral equations diverge, 
forcing the adoption of regularization schemes of some kind. 
The most common regularization procedure consists on 
the inclusion of form-factors, which could, in principle, be 
reinterpreted in terms of self-energy contributions 
of nucleons and mesons \cite{Riska87,Gross92}. 
Nevertheless, the determination from first principles 
of self-energies, due to radiative corrections, would require 
the resolution of coupled Dyson-Schwinger 
equations, again far beyond the present calculations capability. 
Furthermore, the inclusion of form-factors can also be justified 
as an effective way of representing the internal structure of the hadrons and, 
as a consequence, phenomenological form-factors have to be 
considered. 
 
In these situations, where Dirac particles exchange mesons 
with realistic couplings implying the transfer of some quantum numbers,
the quality of the Gross prescription
has not yet been studied systematically. 
Therefore its application in such circunstances has,
at present, only heuristic and pragmatic motivations.
Even so, the Gross prescription has been already applied to 
the 2- and 3-nucleon systems \cite{Gross92,JPinto96,Stadler96}, 
which urges, on one hand, the study of its ability to approximate the exact solutions 
of the Bethe-Salpeter equation with ladder and crossed-ladder series (from 
here on we use exact in this sense) 
 and, on the other, the search for ways to 
systematically improve it.
This study is precisely the goal of the present work. 

The analysis is focused on fourth-order diagrams with incoming and
outgoing particles on-mass-shell,
since the complexity of the calculations is substantial  
and the information obtained is already very rich and relevant. 
These amplitudes
were considered before in references \cite{Gross82,Tjon82}; in the first one
only the static limit was evaluated, in the second one the recent 
NN models were
not considered yet, and the comparison between the Gross prescription and the
exact calculation is presented only for one energy.
In Sec.\ \ref{sec2} a brief 
introduction to the Bethe-Salpeter equation and its 
3-dimensional reductions is given. 
In Sec.\ \ref{sec3} the box and crossed-box pion exchange amplitudes 
are analyzed in detail. 
In Sec.\ \ref{sec4} different form-factors 
are examined and, in particular, new form-factors are suggested 
aiming the resolution of problems inherent to the conventional ones. 
In Sec.\ \ref{sec5} results of applications are discussed and, 
finally, in Sec.\ \ref{sec6} conclusions are presented.  
   
\section{Bethe-Salpeter Equation and three dimensional reductions}
\label{sec2}

Within a covariant field theory framework, the scattering 
amplitude ${\cal M}$ is given by the sum of all the possible interaction 
processes (represented by Feynman diagrams) derived from the Lagrangian 
under consideration. The scattering amplitude of the interaction 
between particles of masses $m$ and $M$ 
is determined by the Bethe-Salpeter (BS) equation \cite{BetheS} 
\begin{eqnarray}   
& &{\cal M}(p,p';P)={\cal V}(p,p';P)+ \label{BS} \\
& & i \int \frac{d^4 k}{(2\pi)^4} {\cal V}(p,k;P) G(k,P) {\cal M}(k,p';P), 
\nonumber
\end{eqnarray}  
where $P$ is the total 4-momentum, and $p$ and $p'$ are respectively 
the initial and final momenta of one of the particles.   
For this equation we use the shorthand notation
\begin{equation}  
{\cal M}={\cal V}+{\cal V}G {\cal M},
\label{shortBS}
\end{equation}  
keeping in mind that the homogeneous term includes a 4-dimensional 
integration.
In the above equation ${\cal V}$ is the interaction kernel
which involves, in principle, all irreducible diagrams, and $G$ is the 
2-particle propagator given by
\begin{equation}  
G(k,P)=G_1(k)G_2(P-k), 
\label{2prop}
\end{equation}  
with the explicit form for Dirac particles  
\begin{eqnarray} 
G_i(k)&=&\frac{-i}{m_i- \not k-i\varepsilon} \label{propagator} \\
&=& \frac{-i (m_i + \not k)}{{m_i}^2- k^2-i\varepsilon},
\nonumber
\end{eqnarray}   
with $m_1=M$, $m_2=m$ and $i=1,2$. 

As mentioned earlier, the exact solution of the BS equation for
general couplings 
is not technically possible at present, and approximations are unavoidable. 
A common approximation consists on restricting ${\cal V}$ 
to a sum of One Boson (or Meson) Exchange diagrams, and is usually 
called the Ladder Approximation. 
There are two main objections to this approximation:
firstly the one body limit (the Klein-Gordon or Dirac equations
with interaction terms) is 
not recovered when one of the masses goes to infinity 
\cite{Gross69,GrossBook}; secondly it has been shown that, in 
general, crossed-ladder diagrams have important contributions 
\cite{Gross69,GrossBook}, and hence should not be neglected.  
 
Alternative ways rely on the replacement of the BS equation 
by the system 
\begin{eqnarray} 
{\cal M}&=&K+ K g {\cal M}  \label{QP1} \\
K&=&{\cal V}+ {\cal V}(G-g) K.
\label{QP2}     
\end{eqnarray} 
which is perfectly equivalent to the BS equation.
In this equations $g$ is a different 2-particle 
propagator and $K$ a new kernel whose iteration 
reads 
\begin{eqnarray}
K&=&{\cal V}+ {\cal V}(G-g) {\cal V} + \label{expancaoK} \\
& &{\cal V}(G-g) {\cal V} (G-g){\cal V} +... 
\nonumber
\end{eqnarray}
A Quasi-Potential equation is obtained through a covariant 
dimensional reduction 
of the Eq.\ (\ref{QP1}). Since there are no first
principle rules to dictate how this reduction should be performed,   
a large ambiguity gives room for many options. 
Usually, different approximations comprise different 
appropriate choices to constraint the energy-component of the 
free 4-momentum in the propagator $g$. If $g$ is a good 
approximation to $G$, then only the first term in Eq.\ (\ref{expancaoK})  
can be kept and a simpler 3-dimensional, but still covariant, 
equation is obtained.    
If needed, and possible, the kernel $K$ can be corrected 
by the inclusion of higher order terms, following Eq.\ (\ref{expancaoK}). 

Examples of QP equations are the Blankenbecler-Sugar \cite{BbS}, 
the Equal-Time \cite{Wallace89} and the Gross \cite{Gross69} equations, 
all satisfying the required one-body limit. 
The Gross equation, hereafter the only considered, results 
from the choice
\begin{eqnarray}
g(k,P)&=&-i 2\pi \frac{ \delta_+(k^2-M^2)}{m^2-(W-k_0)^2-i \varepsilon} 
\Lambda _1(k) \Lambda _2 (P-k) \label{propG} \\
&=&-i 2\pi \frac{ \delta (k_0-E_{\bf k})}
{2E_{\bf k} \left[e_{\bf k}^2-(W-E_{\bf k})^2-i \varepsilon \right]} 
\Lambda _1(k) \Lambda _2 (P-k), \nonumber
\end{eqnarray}  
where 
\begin{eqnarray}
& &W=\sqrt{P^2} \\
& &E_{\bf k}=\sqrt{M^2+{\bf k^2}} \label{Energia} \\
& &e_{\bf k}=\sqrt{m^2+{\bf k^2}} \\ 
& &\delta _+(k^2-M^2)= \delta (k^2-M^2) \theta (k_0)\\ 
& &\Lambda _i(k)= m_i+ \not k.
\end{eqnarray}
The second line of Eq.\ (\ref{propG}) is valid only in the CM 
reference frame. 
As it can be seen from this equation, the Gross choice 
means that the particle with mass $M$ is placed on its mass shell 
in all intermediate states. 

The principal motivation for the Gross equation comes from the cancellation 
theorem \cite{Gross69,GrossBook}, that states that for scalar 
particles in the static limit ($M \to \infty$), the sum of all 
ladder and crossed-ladder diagrams is exactly given by the 
sum of ladder diagrams with the massive particle on its mass shell. 
This is shown to be a result of important cancellations between 
ladder and crossed-ladder amplitudes in all orders. In Fig.\ \ref{cancel} 
this cancellation is illustrated for fourth-order diagrams. In all figures
the cross on a line means that the corresponding particle is on-mass-shell;
we will refer to the diagram with one of the particles on its mass-shell in
the intermediate state as the Gross amplitude. 

For identical interacting particles however, 
equal treatment is required and Eq.\ (\ref{propG}) has to be properly 
symmetrized in order to comply with the Pauli principle. 
This procedure has been implemented recently   
using a system of coupled equations for the half-on-mass-shell amplitudes
\cite{Gross92}.
Each of these equations can be written as follows:
\begin{eqnarray}
& &{\cal M}(p,p',P)={\cal V}(p,p',P) \label{simetricM}\\
& &+i\int \frac{d^4 k}{(2\pi)^4} 
{\cal V}(p,k;P) g(k,P) {\cal M}(k,p';P);
\nonumber
\end{eqnarray}  
here the propagator takes into account an 
adequate symmetrization of the particles in all 
intermediate states having the form: 
\begin{eqnarray}
g(k,P)&=&\frac{1}{2} (2\pi)\Lambda _1(k)G_2(P-k) \delta _+(M^2-k^2)     \\
       & &\frac{1}{2} (2\pi) G_1(k) \Lambda _2(P-k) \delta _+(M^2-(P-k)^2)
\nonumber
\end{eqnarray}
where the equal mass condition ($m=M$) has been used.   
The 3-dimensional reduction is trivially obtained by performing the 
$k_0$ integration using the $\delta _+$ function.  

\section{Box and Crossed-Box amplitudes}
\label{sec3}
 
In this section we consider the
crossed-box diagram and the contributions
to the box 
which are not explicitly contained in the
spectator formalism. The last ones, often denoted by subtracted-box
contributions, are shown in Fig.\ \ref{subtract}.
Together, crossed-box and subtracted-box diagrams
are the lowest order corrections
to the one boson exchange kernel
of the Gross equations (see Eq.\ (\ref{QP2})).
We calculated them
exactly for a variety of kinematic conditions
(energy and scattering angle)
in order to assess their importance relatively
to the Gross amplitude.

We started by deriving the general expressions for
the box and crossed-box diagrams, the
${\cal M}_I$ and ${\cal M}_{II}$ amplitudes
respectively, which are represented in Fig.\ \ref{coordenadas} along with
detailed notation for the kinematic variables in the inner loop. In all
diagrams particle 1 is represented by the lower line and particle 2 by
the upper line.

These amplitudes result from
an effective Lagrangian for 
spin-$1/2$ particles of equal mass $M$,     
interacting through the exchange of a meson 
of mass $\mu$:
\begin{eqnarray}
{\cal L}_{\sigma} &=&
\frac{1}{2} \bar \psi(i\gamma_{\mu} \partial^{\mu}-M)\psi + 
\label{Lsigma}\\
& &\frac{1}{2} (\partial^{\mu} \sigma \partial_{\mu} \sigma -\mu^2 \sigma^2)
-g_{\sigma} \sigma \bar \psi  \psi,   
\nonumber
\end{eqnarray}
for scalar vertices, or 
\begin{eqnarray}
{\cal L}_{\pi}&=&
\frac{1}{2} \bar \psi(i\gamma_{\mu} \partial^{\mu}-M)\psi + 
\frac{1}{2} (\partial^{\mu} {\bf \pi}  \cdot \partial_{\mu} 
{\bf \pi} -\mu^2 {\bf \pi^2}) \nonumber \\
&-&ig_{\pi} \bar \psi 
\left[\lambda \gamma^5+(1-\lambda) \frac{\not q}{2M} \gamma^5 \right]
{\bf \tau \cdot \pi}\psi  
\label{Lpi}
\end{eqnarray} 
for a isovector, pseudoscalar and pseudovector coupling admixture,
with a mixing parameter
$\lambda$ ($0 \leq \lambda \leq 1$).

Application of the Feynman rules gives for the amplitudes:  
\begin{eqnarray}  
{\cal M}_I&=&i \Gamma  \int \frac{d^4 k}{(2\pi)^4}
H_I(p,k,p')[\xi_I^1(k) \otimes \xi_I^2(P-k)]   
\label{BoxMatrix} \\
& & \times D_1(k)D_2(P-k)d_3(p-k)d_4(p'-k)  \nonumber 
\end{eqnarray}
for the box amplitude, and 
\begin{eqnarray}  
{\cal M}_{II}&=&i \Gamma  \int \frac{d^4 k}{(2\pi)^4}
H_{II}(p,k,p')[\xi_{II}^1(k) \otimes \xi_{II}^2(P-k)] 
\label{CrossMatrix} \\
& &\times D_1(p'+p-k)D_2(P-k)d_3(p-k)d_4(p'-k), \nonumber 
\end{eqnarray}
for the crossed-box amplitude.

In the above equation
$\Gamma $ includes the coupling constants and the
isospin operators defined in the Appendix 
(Eqs.\ (\ref{GammaPS}) and (\ref{GammaPV})).
The propagator of spinor $i$, $i=1,2$, has been separated into
a function
$D_i$ (the denominator of the Dirac 
propagator
following the second line of equation
\ref{propagator}) which coincides with the propagator of a
scalar particle with the same mass $M$
\begin{eqnarray} 
D_i(k)=\frac{1}{M^2-k^2-i\varepsilon_i}, 
\end{eqnarray}
and a function carrying
the Dirac structure of the 
spin$-1/2$ particle which, together with the two interaction vertices, defines
the Dirac structure functions $\xi_{\alpha}^i(k)$ ($i=1,2$; $\alpha= I,II$).
Although the explicit form of $\xi_{\alpha}^i(k)$ depends
on the interaction considered, 
it has the general structure 
\begin{equation}
\xi_{\alpha}^i(k)=a_{\alpha}^i(k)+ b_{\alpha}^i(k) \not k .
\label{fi}
\end{equation}
The expressions for $a_{\alpha}^i(k)$ and $b_{\alpha}^i(k)$
are given in the Appendix. The required explicit expressions for 
spinors and phase conventions can be
found in the work of F. Gross et al \cite{Gross92}.

The $d_i$ function stands for the propagator of meson $i$ 
\begin{eqnarray} 
d_i(q)=\frac{1}{\mu^2-q^2-i\varepsilon_i}. 
\end{eqnarray}
Finally, $H_{\alpha}$
are scalar functions involving the 
regularization mechanisms of the vertices, or  
form-factor functions. 
They will be discussed at length in Sec.\ \ref{sec4}. 

We note that in the case of scalar particles, no regularization scheme
is needed, and we may simply take $H_{\alpha}=1$, 
together with 
$a_{\alpha}^i(k) =1$
and  $b_{\alpha}^i(k)=0$, 
($i=1,2$; $\alpha=I,II$). This particular
situation was considered by us as
a test study, meant firstly to check partial analytical and numerical results.

Introducing explicitly the on-mass-shell energy variables
$E_{\bf k}$ defined by Eq.\ (\ref{Energia}) and
\begin{equation}
\omega_{\bf q}=\sqrt{\mu^2+{\bf q}^2},
\end{equation}
and the off-mass-shell energy-component $k_0$ we can write 
\begin{eqnarray}  
& &{\cal M}_{I}=i \Gamma  \int \frac{d^4 k}{(2\pi)^4} 
H_I \;[\xi_I^1 \otimes \xi_I^2] \;h_I({\bf p,k,p'};k_0) \label{MI2} \\
& &{\cal M}_{II}=i \Gamma  \int \frac{d^4 k}{(2\pi)^4} 
H_{II} \;[\xi_{II}^1 \otimes \xi_{II}^2] \;h_{II}({\bf p,k,p'};k_0)
\label{MII2} 
\end{eqnarray}
where part of the integrand function arguments 
have been omitted for simplicity.
The $h_I$ and $h_{II}$ functions are given by:
\begin{eqnarray}
& &
h_I({\bf p,k,p'};k_0)= \\
& &\frac{1}{[E_{\bf k}^2-k_0^2-i\varepsilon _1]
[E_{\bf k}^2-(W-k_0)^2-i\varepsilon _2]} \nonumber \\
& & \times \frac{1}{[\omega _{\bf p-k}^2-(E_{\bf p}-k_0)^2-i\varepsilon _3] 
[\omega _{\bf p'-k}^2-(E_{\bf p}-k_0)^2-i\varepsilon _4]} \nonumber
\end{eqnarray} 
and 
\begin{eqnarray}
& &h_{II}({\bf p,k,p'};k_0)= \\
& &\frac{1}{[E_{\bf p'+p-k}^2-k_0^2-i\varepsilon _1]
[E_{\bf k}^2-(W-k_0)^2-i\varepsilon _2]} \nonumber \\
& &
\times 
\frac{1}
{[\omega _{\bf p-k}^2-(E_{\bf p}-k_0)^2-i\varepsilon _3] 
[\omega _{\bf p'-k}^2-(E_{\bf p}-k_0)^2-i\varepsilon _4]}. \nonumber
\end{eqnarray}  
From the last two equations we can locate the $k_0$ poles,
related with the physical singularities present to
satisfy unitarity. If the form-factor functions $H_{\alpha}$ 
have no singularities, and since each propagator contains two poles
near the real axis,
the $h_{\alpha}$ functions will have eight physical complex
singularities, four in the upper-half complex plane and four in
the lower one. The
$k_0$ integration can be done using the Cauchy 
theorem by closing the integration contour, and for practical reasons we
chose to close the contour integration in
the lower-half complex plane. 
The box amplitude splits into the four following contributions 
\begin{equation}
{\cal M}_I={\cal M}_I^+ + {\cal M}_I^- + {\cal M}_I^{10}+ 
{\cal M}_I^{20}
\label{MI}
\end{equation}
where
\begin{itemize}  
\item
${\cal M}_I^+$ is the residue of the $k_0=E_{\bf k}$ pole, present
in the propagator of  particle 1,  
meaning that this particle is  on-mass-shell
with 4-momentum $(E_{\bf k}, {\bf k})$ (positive energy).
This amplitude is the Gross or
spectator fourth-order amplitude.

\item
${\cal M}_I^-$ is the residue of the $k_0=W+E_{\bf k}$ pole, present
in the propagator of  particle 2,
meaning that this particle is  on-mass-shell 
with 4-momentum $(-E_{\bf k}, -{\bf k})$ (negative energy).

\item
${\cal M}_I^{10}$ is the residue of the $k_0=E_{\bf p}+\omega_{\bf k-p}$
pole, present in the propagator of the meson carrying momentum $p-k$, 
meaning that this meson is 
on-mass-shell with 4-momentum $(\omega_{\bf p-k}, {\bf p-k})$
(positive energy).

\item
${\cal M}_I^{20}$ is the residue of the
$k_0=E_{\bf p}+\omega_{\bf p'-k}$ pole, present in the propagator of the
meson carrying momentum $k-p'$, 
meaning that this other meson is 
on-mass-shell with 4-momentum $(-\omega_{\bf p'-k}, {\bf k-p'})$
(negative energy).
\end{itemize}

Similarly, the crossed-box amplitude decomposes into
\begin{equation}
{\cal M}_{II}={\cal M}_{II}^+ + {\cal M}_{II}^- + {\cal M}_{II}^{10}+ 
{\cal M}_{II}^{20}
\label{MII}
\end{equation}
where
\begin{itemize}  
\item
${\cal M}_{II}^+$ corresponds to the residue of the $k_0=W+E_{\bf p+p'-k}$
pole, meaning that 
particle 1 is on-mass-shell  
with 4-momentum $(-E_{\bf p'+p-k}, {\bf p'+p- k})$
(negative energy).

\item
${\cal M}_{II}^-$  corresponds to the residue of the $k_0=W+E_{\bf k}$ pole,
meaning that particle 2 is on-mass-shell
with 4-momentum $(-E_{\bf k}, -{\bf k})$
(also negative energy).

\item
${\cal M}_{II}^{10}$ and ${\cal M}_{II}^{20}$ have the same  
interpretation as ${\cal M}_{I}^{10}$ and ${\cal M}_{I}^{20}$
have for the box amplitude. 
\end{itemize}

Given the above interpretation we can separate
retardation effects from nucleon negative energy states contributions. 
The retardation effects are obtained by adding 
\begin{equation} 
{\cal M}_I^0={\cal M}_I^{10}+{\cal M}_I^{20}
\label{MI0}
\end{equation} 
from the box diagram to
\begin{equation} 
{\cal M}_{II}^0={\cal M}_{II}^{10}+{\cal M}_{II}^{20}
\label{MII0}
\end{equation} 
from the crossed-box diagram. 
The nucleon negative energy states contributions are given by 
adding ${\cal M}_I^-$ from the box diagram to 
\begin{equation} 
{\cal M}_{II}^{+-}={\cal M}_{II}^{+}+{\cal M}_{II}^{-}
\label{MII+-}
\end{equation} 
from the crossed-box diagram.

It can be proved that if the $H_{\alpha}$ functions  have only real
singularities, the sum of 
${\cal M}_I^0$, ${\cal M}_I^-$
${\cal M}_{II}^0$  and ${\cal M}_{II}^{+-}$ 
is real \cite{Ramalho98}. This result guarantees in particular
that the fourth-order amplitude satisfies unitarity:
the only possible imaginary part comes from the two-nucleon unitarity cut.
Still, the existence of singularities
in the $H_{\alpha}$ functions may yield extra terms to Eqs.\ (\ref{MI})
and (\ref{MII}), other than the physical on-mass-shell particle pole
contributions
discussed. We anticipate here that 
those contributions are in general not negligible.
Our study will henceforth show that the inclusion of crossed-box and
subtracted-box diagrams
in the kernel allows us to be more discriminative in our choices for
the short-range parameterization of the the NN interaction. We will be back to
this discussion in the next sections.

Finally, we note  that the usual procedure for antisymmetrizing
the amplitudes has to be performed. This can be achieved by 
permuting the particles in the final state \cite{Gross92}, as usually.
One obtains for the fourth-order amplitude:
\begin{equation}
{\cal M}^{(4)}=
\frac{1}{2}
\left( {\cal M}_I+\delta \bar {\cal M}_I \right)+
\frac{1}{2}
\left( {\cal M}_{II}+\delta \bar {\cal M}_{II} \right)
\label{M4}
\end{equation}
where $\delta$ is $(-1)^I$ ($I$ being the total isospin of the system) 
and $\bar {\cal M}$ denote the permuted terms, 
obtained by application of the permutation operator to the
final state particles.

Additionally, as discussed in reference \cite{Gross92},
the Pauli
principle imposes an extra symmetrization on
the spectator formalism amplitudes: this symmetrization is
intrinsic to the intermediate states, where only one of the particles is 
on-mass-shell. As a consequence, 
the fourth-order diagrams resulting from the Gross equations are 
not only the diagrams with 
particle 1 on-mass-shell in the intermediate state, 
amplitude ${\cal M}_I^+$, 
but also the diagrams where particle 
2 is on-mass-shell in the intermediate state,
amplitude ${\cal M}_I^{+\,\prime}$, as can be seen from Fig.\ \ref{4ordem}:
\begin{eqnarray}
{\cal M}^{(4)}_{Gross}&=&
\frac{1}{4}{\cal M}_I^+ +\frac{\delta}{4}\bar {\cal M}_I^+ +\\
& &\frac{1}{4}{\cal M}_I^{+\,\prime} +
\frac{\delta}{4}\bar {\cal M}_I^{+\,\prime}. \nonumber
\end{eqnarray}
The coefficients affecting the several sub-amplitudes
force the two particles to be {\it equally} on-mass-shell in the 
intermediate state, which is sufficient for
the total amplitude to satisfy the
Pauli Principle.
 
\section{Strong form-factors}
\label{sec4}

Nucleons are not elementary particles and consequently
their hadronic structure has to be taken into account.
Mathematically, such structure provides the necessary regularization
of the integrals for the high order loops, needed to build in
the analytical structure of a bound state wavefunction, or of
a scattering transition matrix.
Unfortunately, it is not yet possible to obtain the
description of the nucleon structure  directly from the QCD Lagrangian, or
even from effective models inspired on constituent quark structure.
The compositeness of the nucleons is therefore currently described  
by means of form-factors which have to be phenomenologically fixed.
Yet, it has been shown by Gross and Riska \cite{Riska87} that
it is possible to describe consistently
the electromagnetic 
interaction with nucleons, within a
relativistic and manifestly covariant framework, with
phenomenological strong form-factors for the meson-nucleon vertices,
present in the 
current conservation relations following the Ward-Takahashi
identities.

In non relativistic 
applications, analytic dipolar (or even higher
power) functions of the 3-momentum are used as form-factors.
In relativistic applications 
the covariance requirement motivates the use of form-factors of the type:
\begin{equation}
f(q^2)=\frac{\Lambda^2}{\Lambda ^2-q^2}
\end{equation}
where $q^2=q_0^2-{\bf q^2}$ substitutes ${\bf q^2}$ 
in the non-relativistic form-factors.  

The introduction of the 
strong form-factors can be interpreted as vertex-dressing, 
and we take them to have a factorized separable form, as done in
reference \cite{Gross92}:
\begin{equation}
F_m(p,p')=f_m(q^2)f_N(p^2)f_N(p^{\prime\,2})
\end{equation}
where $f_N$ and $f_m$ stand for the nucleon and meson form-factors 
respectively,
$p$ and $p'$ are the nucleon 4-momenta and $q=p'-p$ the meson 4-momentum. 
As a consequence, the $H_{\alpha}$ function, introduced in the 
Sec.\ \ref{sec3}, includes from each vertex  
two nucleon form-factors
$f_N$ and one meson form-factor $f_m$.
The function $H_{\alpha}$ is therefore a product of eight
form-factor functions, since the form-factors of on-mass-shell
nucleons are normalized to one.

Alternatively, the dressing can be associated with self-energy 
corrections to the Dirac and meson propagators (radiative corrections). 
Following these lines, a product of two strong form-factors $f_m(q^2)$, each
coming from a different vertex, can be incorporated in the renormalization of 
the meson propagator according to reference \cite{Riska87}:
\begin{equation}
\frac{f_m^2(q^2)}{\mu^2-q^2}
=\frac{1}{\mu^2-q^2+\Pi_m (q^2)}
\end{equation}
being the meson self-energy becomes fixed by

\begin{equation}
\Pi_m(q^2)=\left[\frac{1}{f_m^2(q^2)}-1\right]
(\mu^2- q^2).
\label{eqPI}
\end{equation}

Analog equations for the nucleon Dirac propagator and self-energy
can be written \cite{Gross92}:
\begin{equation}
\frac{f_N^2(p^2)}{M-\not p}
=\frac{1}{M-\not p+ \Sigma (p)}
\end{equation}
being the nucleon self-energy becomes fixed by 

\begin{equation}
\Sigma(p)=\left[\frac{1}{f_N^2(p^2)} -1\right]
(M-\not p).
\label{eqSigma}
\end{equation}

In the most recent works which applied the Gross equation
\cite{Stadler96}
the form-factors were assumed to have the expression:
\begin{eqnarray}
& &f_N(p^2)=
{{(\Lambda _N^2-M^2)^2} 
\over {(\Lambda _N^2-M^2)^2+(M^2-p^2)^2}} \label{fN1} \\
& &f _{m}(q^2)=
{{(\Lambda _m ^2-\mu ^2)^2+\Lambda _m ^4} 
\over{(\Lambda _m ^2-q^2)^2+\Lambda _m ^4}} \label{fPI1}
\end{eqnarray}
The two parameters $\Lambda _N$ and
$\Lambda _m$ are the nucleon and meson cut-offs respectively and 
are fixed, along with the couplings and meson masses, through a fit
to the nucleon-nucleon scattering data. 

These choices imply that the {\it effective} 
nucleon propagator for those models is the product 
\begin{eqnarray}
& &\frac{f_N^2(p^2)}{M-\not p}
=\frac{1}{M-\not p} \times \label{facturiza} \\ 
&\times&\left[\frac{\tilde \Lambda_N^2}{M^2+i\tilde \Lambda_N^2-p^2} \right]^2
\left[\frac{\tilde \Lambda_N^2}{M^2-i\tilde \Lambda_N^2-p^2} \right]^2
\nonumber
\end{eqnarray}
with $\tilde \Lambda _N^2= \Lambda_N^2-M^2$. A similar expression 
can be written for the meson effective propagator.  
The first factor on the r.h.s. of Eq.\ (\ref{facturiza}) 
is the {\it non renormalized} 
propagator, while the second and third factors are effectively   
propagators of resonances
of width $\tilde \Lambda_N^2$. Such widths have ad-hoc energy and 3-momentum
dependencies, in particular they are constant in the entire
energy and momentum range, with the consequence of inducing spurious
imaginary contributions in the scattering amplitude.
We can interpret these complex mass terms  
as inelasticity sources associated with nucleon 
or meson excitations, which nevertheless are non physical,
namely because they open channels at energies below the 
pion production threshold.

As we will see in the next section, the spurious contributions may be small 
under some conditions, e.g. for light meson scalar exchange, 
but they have sizeable effects for a 
pion exchange with pseudovector coupling.
 
The pathology of the form-factor that we are discussing
manifests only when we use it in the {\it full} 
Bethe-Salpeter equation, that is to say, when we  
treat the energy-component as a totally independent variable. 
For the Gross reduction, where these form-factors were used
and calibrated by  the data,  
the spurious components are not evident: the prescription
for the $k_0$ variable  constrains its range such that the 
non-physical complex singularities are irrelevant. In other words,
the spectator equation is insensitive to the existence of the complex
poles of the form-factors. Nevertheless, the study of the validity
of the Gross prescription, as well as the introduction of possible required corrections
to the kernel of the integral equation, need to incorporate crossed-box
and subtracted-box diagrams which may be crucially affected by
the form-factors poles.

In order to solve the problem we explored
alternative form-factor functions. Since the form-factors are
essentially needed to cut the 3-momentum range of the integrals,
we made the choice of not allowing the regularization function to depend
on the relative energy variable. However, in order to still satisfy
covariance, the new form-factors 
were built by giving to the square of the 
3-momentum  ${\bf k}$ a covariant form. More specifically,
in an arbitrary frame
where the total two nucleon momentum is $P$, ${\bf k}^2$
is replaced by the Lorentz invariant
\begin{equation}
\bar k^2=\frac{(P \cdot k)^2}{P^2}-k^2,
\label{kCM}
\end{equation}
with $\bar k^2={\bf k}^2$ in the CM frame. $\bar k$ stands for
the square root of $\bar k^2$. 
It becomes clear from this equation that the price 
to be paid for this new choice, which does not consider
the relative energy dependence explicitly, is to have form-factors depending 
on the on-mass-shell momentum of the external particles, or, in other words,
on the total momentum  $P$ of the 2 nucleon system. The  complexities
implied by this choice are only relevant, but with the possibility
of being actually handled in practice, for the 3-nucleon applications,
where the 2-nucleon subsystem has to be treated in the 3-nucleon system
CM frame. They will also imply a new version of the work on the
electromagnetic currents of reference \cite{Riska87}, 
since in this reference the
nucleon-nucleon interaction does not depend on 
the total momentum of the system.

Examples of this type of form-factor choices 
can be found, for instance, in the work of the Nijmegen \cite{Rijken} and
Bonn \cite{Bonn} groups that used
\begin{equation}
f_m({\bar q^2})=e^{-\frac{\bar q^2}{2 \Lambda ^2}}
\end{equation}
and
\begin{equation}
f_m({\bar q^2})=\left[\frac{\Lambda^2}
{\Lambda^2+{\bar q^2}} \right]^n.
\label{multipole}
\end{equation}
respectively, which were already written in terms of the 
Lorentz invariant $\bar q^2$.

For our calculations we took two different models for the meson 
form-factor functions. Both choices satisfy the criterion of leaving out
the $k_0$ dependence from those functions. The first choice corresponds to 
Eq.\ (\ref{multipole}) with $n=1$.
For the second choice, we selected
a functional form which would furthermore not alter the results of
the calculations
already done within the spectator formalism, preserving namely the
description of the two nucleon scattering observables. Explicitly, we used 
\begin{eqnarray}
f_m({\bar p^2}, {\bar p^{\prime\,2}},{\bar q^2})&=&
\left.
\frac{(\Lambda^2_m-\mu^2)^2+\Lambda^4_m}
{ (\Lambda^2_m-q^2)^2 
+\Lambda^4_m} 
\right|_{k_0=E_{\bar k}} \nonumber \\
& =&\frac{(\Lambda^2_m-\mu^2)^2+\Lambda^4_m}
{ \left[\Lambda^2_m+{\bar q^2}-(E_{\bar p}-E_{\bar p'})^2 \right]^2
+\Lambda^4_m}
\label{fPI3}
\end{eqnarray}
The $k_0=E_{\bar k}$ condition is necessary in
order to keep the results for the ${\cal M}_I^+$ amplitude, that are
part of the calculations 
in references \cite{Gross92,Stadler96}. 
This is the reason why in the denominator
a "retardation-like" term $E_{\bar p}-E_{\bar p'} $ appears , 
even though it may seem
arbitrary at first thought. We may advance at this point 
that this term has very
little importance: the results with and without it differ by at most 10\%,
with the exception of the region where the amplitudes vanish in the $I=1$
channel.

A new version of the nucleon form-factors $f_N$ is also required.
We again made them dependent only on the relative 3-momentum which,
together with the manifest covariance requirement,
made them depending
on the total 2 nucleon 4-momentum (Eq.\ (\ref{kCM})). 
Explicitly, we took the general nucleon-form-factor used in reference 
\cite{Stadler96} calculated as for the ${\cal M}_I^+$ amplitude,
which reads
\begin{eqnarray} 
f_N({\bar k^2};{\bar p^2})=
\left[
\left.
\frac{\tilde \Lambda_N^4}
{\tilde \Lambda_N^4 +(P-k)^2} 
\right|_{k_0=E_{\bar k}}
\right]^{1/2} \nonumber \\
\left[
\frac{\tilde \Lambda_N^4}{\tilde \Lambda_N^4 +W^2(W-2E_{\bar k})^2} 
\right]^{1/2}.
\label{fN2}
\end{eqnarray}
In this equation $W=2E_{\bar p}$ is the invariant mass, ${\bar p^2}$ is
defined from the 3-momentum of the initial and final particles and 
${\bar k^2}$ is defined from the off-mass-shell particle 3-momentum. 

The introduction of these new form-factor functions allowed us
to eliminate the imaginary spurious contributions to the box and crossed-box
diagrams, and successfully construct an amplitude
satisfying the elastic cut unitarity requirement. This conclusion will
be accounted for in detailed and quantitative terms in the next section.

\section{Results and Discussion}
\label{sec5}

In order to describe the scattering process 
between two Dirac interacting particles we need to specify,
in addition to the energy and scattering angle,
the initial and final isospin and helicity (or spin) states. 
In general we have five independent helicity channels from a
total of sixteen possible.
We present here only the results for the matrix element
corresponding to the $++\to ++$ transition, since similar
conclusions can be drawn for the other channels.

If the form-factors have no singularities, we can use  
the decomposition of Eqs.\ (\ref{MI}) and (\ref{MII}) to
compute the amplitudes. 
This is not the case, however, of
the form-factors of Eqs.\ (\ref{fN1}) and (\ref{fPI1})
whose singularity structure implies the inclusion 
of the corresponding residues. 

\subsection{Scalar exchange}

As a reference calculation (an important check to our analytical
calculations and numerical results) we 
considered the simplest case of
scalar exchange between scalar identical particles of mass $M$ (equal masses),
where no form-factors are needed. 
We used the parameters 
\begin{eqnarray}
& &\frac{g^2}{4\pi M^2}=0.5 \nonumber \\
& &\mu=M/7. \nonumber 
\end{eqnarray}
The results are shown in Fig.\ \ref{scalar=m} for
three different kinetic energies in the
lab reference frame. The curves refer
to the box and crossed-box amplitudes, their sum and 
the spectator amplitude. For comparison we also include
the result for the amplitude corresponding to the
Blankenbecler-Sugar quasi-potential equation.

The figure clearly shows the
cancellation between parts of the box and the
crossed-box amplitudes, which has been shown
previously to occur only in the static limit
\cite{Gross69,GrossBook}.
This conclusion
is interesting, and in the meantime 
path integral techniques
have been used to verify its validity 
for all orders \cite{Tjon}.

Next we considered the exchange of a scalar
meson between nucleons, which 
forces us to consider the $I=0$
and $I=1$ isospin NN channels. Form-factors of 
Eqs.\ (\ref{fN1}) and (\ref{fPI1}) were included,
and the parameters were taken from the models used in the three-nucleon
calculations \cite{Stadler96,Private}, which read 
\begin{eqnarray}
& & \frac{g_{\sigma}^2}{4\pi}=4.82 \nonumber \\
& & \mu =m_{\sigma}= 498 \mbox{ MeV}\nonumber \\
& & \Lambda _{\sigma}= \Lambda_m= 1141 \mbox{ MeV}\nonumber \\
& & \Lambda_N= 1862 \mbox{ MeV}.\nonumber
\end{eqnarray}
We note that our calculation does not 
include the general off-shell couplings of the original 
model \cite{Stadler96}.

The results for laboratory kinetic energy $T_{lab}$ 
of 100, 200, and 300 MeV are shown in
Figs.\ \ref{Sigma0} and \ref{Sigma1}, for the
$I=0$ and $I=1$ cases respectively. 
One concludes that
the real part of the sum of the box and crossed-box amplitudes
is well differs from the Gross amplitude, being
the relative deviation of the order 
of 15\% for both isospin channels, at 100 MeV, and
30\% at 200 MeV. For 
300 MeV these deviations become 50\% for $I=0$ and 
70\% for $I=1$. This quantifies how the approximation
deteriorates with increasing energy. 
It is interesting to add that in general the fourth
order $\sigma$ exchange contributions 
are 10\% of the second order (one $\sigma$ exchange).

The quality of the Gross prescription is slightly improved 
for a lighter exchanged meson, as shown in Figs.\ \ref{pion0} and  
\ref{pion1},
where the results were obtained with:
\begin{eqnarray}
& &\frac{g_{\pi}^2}{4\pi}=1 \nonumber \\
& &\mu =m_{\pi}=138 \mbox{ MeV}  \nonumber \\
& &\Lambda _m=\Lambda _{\pi}= 2109 \mbox{ MeV} \nonumber
\end{eqnarray}
Since this exchange does not correspond to the pseudovector pion exchange,
the value of the coupling 
constant was arbitrarily chosen by the convenience. 
This situation has been considered to study 
how the quality of the Gross prescription depends on the 
meson mass.  
From the same figures we can conclude also that while at 100 MeV
the crossed-box amplitudes are very small, they become
important at higher energies and cannot be neglected, as 
assumed by the ladder approximation.  
As for the retardation effects we can say that the
contributions from the meson poles in the box
and crossed-box tend to cancel each other. We may then
conclude that the ladder approximation overestimates
meson retardation effects.
 
Note that we showed only the real part of the 
fourth-order amplitudes. As discussed in the Sec.\ \ref{sec3}, 
the only  imaginary contribution comes from the spectator 
amplitude ${\cal M}_I^+$ when the form-factors have no singularities. 
This is not obviously the case of the form-factors of Eqs.\ (\ref{fN1}) 
and (\ref{fPI1}), but fortunately the additional spurious 
residues contributions are small and mean less than a 2\% contribution. 
So in this case the unitarity violation is very small.

\subsection{PV coupling}
\label{pvold}

Next we evaluated the box and crossed-box amplitudes 
for the $\pi$NN coupling of Eq.\ (\ref{Lpi}) restricted to 
PV coupling ($\lambda =0$). We also used the
form-factors introduced in Eqs.\ (\ref{fN1}) and (\ref{fPI1}). 
The pion mass and cut-off parameters were given above and  
the coupling constant taken from the references 
\cite{Stadler96,Private}
$$
\frac{g_{\pi}^2}{4\pi}=13.34. 
$$

The allowed exchange of isospin introduces weighting factors for the box and
crossed-box that are not the same for the
two different isospin channels, resulting for the
fourth-order amplitude
\begin{equation}
{\cal M}^{(4)}=3({\cal M}_I+{\cal M}_{II})
-2({\cal M}_I-{\cal M}_{II}) \tau_1 \cdot \tau_2.
\end{equation}
The sign  difference in the coefficients
affecting the
crossed-box contribution in the
isoscalar and isovector parts, makes
difficult the simultaneous representation of
the exact amplitudes, in the two isospin channels,
by the spectator-on-mass-shell
approximation.

The real and imaginary parts of the amplitudes for $T_{lab}=100$ MeV 
are represented in Fig.\ \ref{PVW18}. The first
observation to be made is that their
magnitudes, specially for the case of
the imaginary part, are abnormally 
large: the OBE contributions corresponding to exchanged
heavy mesons of the same model 
are $\sim$ 100 $GeV^{-2}$. 
The second observation is that contrarily to the scalar 
coupling case the imaginary contribution of the crossed-box amplitude, 
that should be exactly zero, 
does not vanish, being two orders of magnitude larger than the box 
imaginary part. These 
conclusions are true for both isospin  channels. 
As mentioned earlier, the reason for these results lies in the complex
singularities present in the form-factors of 
(\ref{fN1}) and (\ref{fPI1}),
which clearly violate the unitarity condition. 

Small changes on the 
cut-off parameters $\Lambda_{\pi}$ 
modify the final results, but do not change 
the orders of magnitude obtained. The figures also
show the dominance of the
crossed-box amplitude justified
by a cumbersome 
combination of the presence of 
$\gamma^5$ in the coupling and the $f_{\pi}(q^2)$ 
form-factor. As can be seen from Fig.\ \ref{plotFPI}, 
this form-factor peaks in the region $q^2 \geq 4 M^2$, 
where the negative-energy nucleon poles,
defining the ${\cal M}_{II}^+$ and ${\cal M}_{II}^-$  contributions, occur:
consequently these amplitudes are largely (and artificially) enhanced. 
Contrarily, in 
the box amplitude, the negative energy-state contribution
included in ${\cal M}_{I}^-$ is strongly suppressed by the nucleon 
$f_N(k^2)$ form-factor, since one of the nucleons is very much
off-mass-shell.
Furthermore, since the region involved in 
${\cal M}_{I}^+$ (the Gross amplitude)
is $q^2\le 0$, this amplitude is not crucially affected by the pronounced peak
of the meson form-factor. The spectator equation solutions are then not
affected by the spurious singularities for the form-factors, as announced in
Sec.\ \ref{sec4}.

\subsection{PV coupling -- new form-factors}
\label{pvnew} 

As a consequence of the above results, we tailored alternative choices 
for the meson and nucleon form-factors. We tested two different models:
model A combines 
the pion form-factor of Eq.\ (\ref{fPI3}) with the nucleon form-factor
of Eq.\ (\ref{fN2}); in model B 
the pion form-factor of Eq.\ (\ref{multipole}) with $n=1$ is
used together with the nucleon form-factor  of Eq.\ (\ref{fN2}).
Since those form-factors have no singularities in the $k_0$
variable, the only terms to be considered in the amplitudes 
calculation are given by
the decompositions of Eqs.\ (\ref{MI}) and (\ref{MII}). 

As explained in Sec.\ \ref{sec4}, model A preserves the spectator amplitude 
${\cal M}_I^+$. Nevertheless it changes crucially the remaining
amplitudes relatively 
the models of reference \cite{Stadler96,Private}. Models A and B
give in general very similar results as can be seen from 
Figs.\ \ref{ABcomp1} and \ref{ABcomp2}, with the exception of the Gross 
amplitude in the $I=1$ channel. Anyhow, since equal conclusions can be 
drawn from 
both models, we present only the final results for model A.

From Figs.\ \ref{ModeloA0} and \ref{ModeloA1} it is clear that the exact and
Gross amplitudes do
differ much less than in the model considered in the
previous sub-section. They are now
of the same order of magnitude, being the
difference between them of the same order of   
OBE scalar exchange amplitude ($\sim$ 100 $GeV^{-2}$), 
which gives the hope that the spectator-on-mass-shell
prescription can be easily improved through the inclusion
of OBE-type terms in the kernel. From the same figures
we conclude that the box
amplitude is dominant for the $I=0$ channel, although 
the crossed-box amplitude is by no means negligible for the $I=1$ channel.

Having found a way of preserving the unitary condition, 
we can now analyze relativistic contributions such as 
retardation, given by ${\cal M}^0_I$ (box) and ${\cal M}^0_{II}$
(crossed-box), and nucleon negative-energy states, given by ${\cal M}^-_I$
(box) and ${\cal M}^{+-}_{II}$ (crossed-box).

In Figs.\ \ref{BoxAnI=0} and \ref{BoxAnI=1} we separately show
those effects for the box amplitude, 
and in the Fig.\ \ref{CrossAnI=0} and 
\ref{CrossAnI=1} for the crossed-box amplitude.
For the box amplitude the meson pole contributions are always 
important, and with opposite sign to the total amplitude. We conclude that
in this case retardation effects
cannot be neglected. The nucleon negative-energy state contributions are now
very much suppressed relatively to the case of Sec.\ \ref{pvold},
and typically of the order of $\sim$ 15\%.  
For the crossed-box amplitude, the meson poles and negative energy 
contributions 
are smaller than for the box, and 
have a partial important cancelation between them, giving rise
to a net small result.   
This last cancellation decreases slowly with increasing energy.

\section{Conclusions}
\label{sec6}

Recently, methods based on path-integral techniques allowed field-theory
calculations, including ladder and crossed-ladder series,
for scalar particles to be performed to all orders \cite{Tjon}. 
When the status of the calculations will reach 
the point of dealing with Dirac 
particles and
general couplings, the comparison of their results with the ones obtained with
different Quasi-Potential formalisms will clarify the unsettled point of
selecting between different covariant formulations, for the
description of nuclear systems.

In the meantime we decided to perform a study to quantify how well 
the amplitudes,
calculated within the spectator-on-mass-shell formalism, represent 
the exact ones when
Dirac particles interact through boson-exchange, in particular 
pion-exchange. This will 
eventually orient us towards possible required corrections, 
which may get increasingly
important at higher energies, namely above the pion production threshold.
We considered here two-pion exchange contributions not
included in the spectator-on-mass-shell formalism. We aimed 
to check whether they can be
included in a kernel restricted to have the OBE form, such 
that they can more easily
be incorporated in current calculations and present-day computer codes.

The main conclusions of our work are:

\begin{enumerate}

\item The form-factors of the effective NN models calibrated with
the spectator equations
can not be used to investigate the quality of the 
spectator amplitudes, since the
corresponding crossed-box and subtracted-box amplitudes, 
required for this study,
strongly violate unitarity.

\item For the same reason the above NN models 
are not adequate for use in extensions of the kernel that 
include crossed-box and subtracted-box terms.

\item With appropriate choices of the form-factors, meaning no 
unitary violation,
the difference between the full and Gross fourth-order amplitudes 
is of the order of magnitude of typical 
OBE scalar exchange amplitudes, raising the expectation that this difference 
may be parameterized by
this type of exchange terms. The study of possible improvements of the
spectator-on-mass-shell prescription, through the inclusion of 
OBE-type terms in the
integral equation kernel, is planned for the near future.

\item There is a crucial (energy-dependent) 
interplay between retardation effects and nucleon
negative-energy state contributions in the box and crossed-box diagrams,
which does not allow the ladder approximation to work well as a representation
of the exact Bethe-Salpeter series (ladder plus crossed-ladder).
This is consistent with the findings of Nieuwenhuis et al. \cite{Tjon}.

\end{enumerate}

\bigskip
\begin{center}
{\bf ACKNOWLEDGMENTS}
\end{center}

The authors wish to thank Franz Gross for very helpful 
discussions and suggestions, and J. A. Tjon for his useful advice.
They also thank A. Stadler for 
many discussions and for having explained
the details of the NN models.
This work was performed under the grants PRAXIS XXI 2/2.1/Fis/223/94 and
PRAXIS XXI BD/9450/96.

\appendix

\section{Coupling parameters coefficients} 
\label{ap1}

\subsection{Scalar vertex}

\begin{equation}
\Gamma=g_{\sigma}^4
\label{GammaPS}
\end{equation}

\begin{eqnarray}
& &a_I^1=M \\
& &b_I^1=1
\end{eqnarray}

\begin{eqnarray}
& &a_I^2= a_{II}^2=M \\
& &b_I^2=b_{II}^2=1
\end{eqnarray}

\begin{eqnarray}
& &a_{II}^1=3M \\
& &b_{II}^1=-1
\end{eqnarray}

\subsection{Pseudovector-pseudoscalar mixture}

\begin{equation}
\Gamma=g_{\pi}^4
\left\{ 
\begin{array}{cc}
{3-2 \tau_1 \cdot \tau_2} & \;\;{\mbox{Box}} \\
{3+2 \tau_1 \cdot \tau_2} & \;\;{\mbox{Crossed-Box}}  
\end{array}
\right.
\label{GammaPV}
\end{equation}

\begin{eqnarray}
& &a_I^1=
{\lambda ^2M+\lambda (1-\lambda )
{{M^2+ k^2} \over M}+(1-\lambda )^2{{M^2+3 k^2} \over {4M}}} \\
& &b_I^1 ={\lambda ^2+2\lambda (1-\lambda )+
(1-\lambda )^2{{3M^2+(P-k)^2} \over {4M^2}}} 
\end{eqnarray}

\begin{eqnarray}
a_{I}^2=a_{II}^2&=&
\lambda ^2 M+\lambda (1-\lambda )
\frac{M^2+(P-k)^2}{M} \\ 
& & +(1-\lambda)^2 \frac{M^2+3(P-k)^2}{4M} \nonumber 
\end{eqnarray}

\begin{eqnarray}
b_{I}^2=b_{II}^2&=&
 -\lambda ^2+2\lambda (1-\lambda) \\
& &+(1-\lambda )^2 \frac{3M^2+(P-k)^2}{4M^2}
\nonumber 
\end{eqnarray}

\begin{eqnarray}
a_{II}^1&=&-\lambda ^2M+\lambda (1-\lambda ) 
\frac{3M^2-(p'+p-k)^2}{M} \\
& &+(1-\lambda )^2 \frac{5M^2-(p'+p-k)^2}{4M} 
\nonumber
\end{eqnarray}

\begin{eqnarray}
b_{II}^1&=&
\lambda ^2+2\lambda (1-\lambda ) \\
&+&(1-\lambda )^2 \frac{3M^2+(p'+p-k)^2}{4M^2}
\nonumber
\end{eqnarray}

\subsection{Form-factor structure}

Box amplitude:

\begin{eqnarray}
H_I(p,k,p')&=&[f_N(k^2)]^2[f_N((P-k)^2)]^2 \\
& &[f_m((p-k)^2)]^2[f_m((p'-k)^2]^2
\nonumber
\end{eqnarray}

Crossed-box amplitude:

\begin{eqnarray}
H_{II}(p,k,p')&=&[f_N((p'+p-k)^2)]^2[f_N((P-k)^2)]^2 \\
& &[f_m((p-k)^2)]^2[f_m((p'-k)^2)]^2
\nonumber
\end{eqnarray}

\newpage

\begin{figure}
\caption{ Fourth-order exact cancellation for scalar  
particles in the static limit. 
The thick line indicates the heavier particle, and  
the cross on the propagator line means 
that the particle is on-mass-shell. The diagram with the cross defines
the Gross amplitude.} 
\label{cancel}
\end{figure}

\begin{figure}
\caption{Subtracted-box diagrammatic definition.}
\label{subtract}
\end{figure}

\begin{figure}
\caption{Box (${\cal M}_I$) and crossed-box (${\cal M}_{II}$)
diagrams and notation for the 
kinematical variables. The sum of the two diagrams defines the exact
fourth-order amplitude.}
\label{coordenadas}
\end{figure}

\begin{figure}
\caption{ Symmetrized Gross amplitude for identical particles. 
The initial and final particles are on-mass-shell 
(cross symbol). 
The phase $\delta$ is $1$ for scalar particles, and 
$(-1)^I$ for nucleons, where $I$ is the total isospin.}
\label{4ordem}
\end{figure}

\begin{figure}
\caption{Real part of the fourth-order amplitudes for equal
scalar particles of mass $M$; 
the exchange scalar meson has a mass of $M/7$.
The solid line corresponds 
to the exact result,
the dashed line to the box amplitude, the long-dashed line
to the crossed-box and the dotted line to the Gross amplitude.}
\label{scalar=m}
\end{figure}

\begin{figure}
\caption{Real part of the fourth-order amplitudes
for interacting nucleons in the isospin $I=0$ channel;
the exchange scalar meson has a mass of 498 MeV.
The curves have the same
meaning as in Fig.\ \ref{scalar=m}.}
\label{Sigma0}
\end{figure}

\begin{figure}
\caption{Real part of the fourth-order amplitudes
for interacting nucleons in the isospin $I=1$
channel;
the exchange scalar meson has a mass of 498 MeV.
The curves have the same
meaning as in Fig.\ \ref{scalar=m}.}
\label{Sigma1}
\end{figure}

\begin{figure}
\caption{Real part of the fourth-order
amplitudes for interacting nucleons 
in the isospin $I=0$ channel;
the exchange scalar meson has a mass of 138 MeV.
The curves have the same
meaning as in Fig.\ \ref{scalar=m}.}
\label{pion0}
\end{figure}

\begin{figure}
\caption{Real part of the fourth-order
amplitudes for interacting nucleons in the isospin $I=1$
channel;
the exchange scalar meson has a mass of 138 MeV.
The curves have the same
meaning as in Fig.\ \ref{scalar=m}.}
\label{pion1}
\end{figure}

\begin{figure}
\caption{Real and imaginary parts of the fourth-order
amplitudes for 
interacting nucleons in the
isospin $I=0,1$ channels;
the exchange pseudovector meson has a mass of 138 MeV.
The curves have the 
same meaning as in Fig.\ \ref{scalar=m} and
correspond to a kinetic energy of 
100 MeV.  
We can see a dominance of the crossed-box amplitude, 
which is almost coincident with the full result in some cases. 
Also the Box and Gross lines are almost coincident in same cases.}
\label{PVW18}
\end{figure}

\begin{figure}
\caption{Pion form-factor of references
\protect \cite{Stadler96,Private}.}
\label{plotFPI}
\end{figure}

\begin{figure}
\caption{Real part of the Gross amplitude for 
interacting nucleons in the isospin $I=0,1$ channels;
the exchange pseudovector meson has a mass of 138 MeV.
Comparison between models A and B
for three different kinetic energies.}
\label{ABcomp1}
\end{figure}

\begin{figure}
\caption{Real part of the box and crossed-box amplitudes for 
interacting nucleons in the isospin $I=0,1$ channels;
the exchange pseudovector meson has a mass of 138 MeV.
Comparison between models A and B for three different kinetic energies.}
\label{ABcomp2}
\end{figure}

\begin{figure}
\caption{Real part of the fourth-order amplitudes for interacting nucleons 
in the isospin $I=0$ channel;
the exchange pseudovector meson has a mass of 138 MeV. 
The curves have the same meaning as in
Fig.\ \ref{scalar=m}. Model A has been
considered.}
\label{ModeloA0}
\end{figure}

\begin{figure}
\caption{Real part of the fourth-order
amplitudes for interacting nucleons 
in the isospin $I=1$ channel;
the exchange pseudovector meson has a mass of 138 MeV. 
The curves have the same meaning as
in Fig.\ \ref{scalar=m}. Model A has been
considered.}
\label{ModeloA1}
\end{figure}

\begin{figure}
\caption{Retardation and nucleon negative-energy contributions
for the box amplitude, in the $I=0$ isospin channel, for
a pseudovector exchanged meson of mass 138 MeV.
Model A has been
considered. The full line
is the total box amplitude, the long-dashed
line is the ${\cal M}_I^-$ amplitude and the dashed line the ${\cal M}_I^0$
amplitude}
\label{BoxAnI=0}
\end{figure}

\begin{figure}
\caption{Retardation and nucleon negative-energy contributions
for the box amplitude, in the $I=1$ isospin channel, for
a pseudovector exchanged meson of mass 138 MeV.
Model A has been
considered.
The full line is the total box amplitude, the long-dashed
line is the ${\cal M}_I^-$ amplitude and the dashed line the ${\cal M}_I^0$
amplitude.}
\label{BoxAnI=1}
\end{figure}

\begin{figure}
\caption{Retardation and nucleon negative-energy contributions
for the crossed-box amplitude, in the $I=0$ isospin channel, for
a pseudovector exchanged meson of mass 138 MeV. Model A has been
considered. The full line is the total crossed-box amplitude, the long-dashed
line is the sum of the ${\cal M}_{II}^+$ and ${\cal M}_{II}^-$ amplitudes
and the dashed line the ${\cal M}_{II}^0$ amplitude.}
\label{CrossAnI=0}
\end{figure}

\begin{figure}
\caption{Retardation and nucleon negative-energy contributions
for the crossed-box amplitude, in the $I=1$ isospin channel, for
a pseudovector exchanged meson of mass 138 MeV.
Model A has been
considered. The full line is the total crossed-box amplitude,
the long-dashed
line is the sum of the ${\cal M}_{II}^+$ and ${\cal M}_{II}^-$ amplitudes
and the dashed line the ${\cal M}_{II}^0$ amplitude.}
\label{CrossAnI=1}
\end{figure}

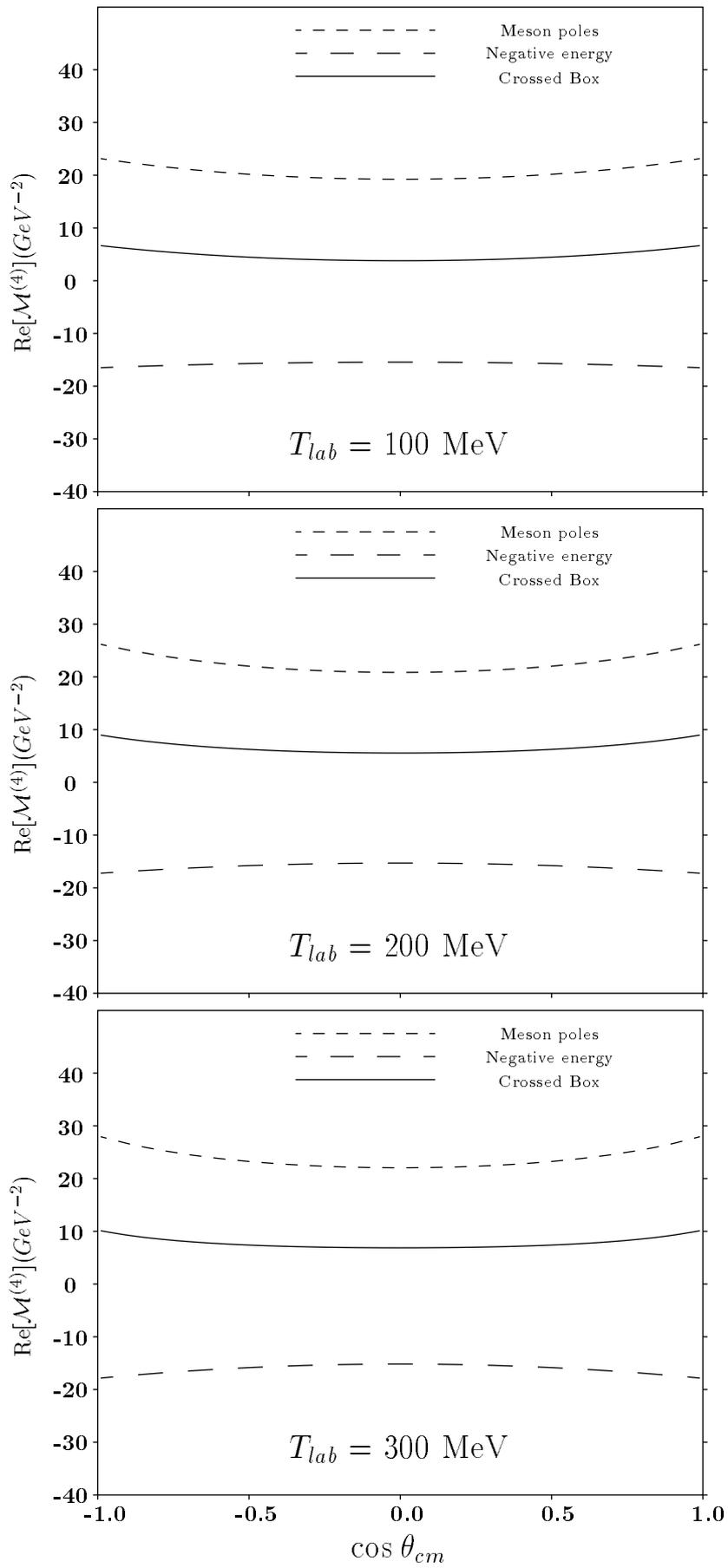
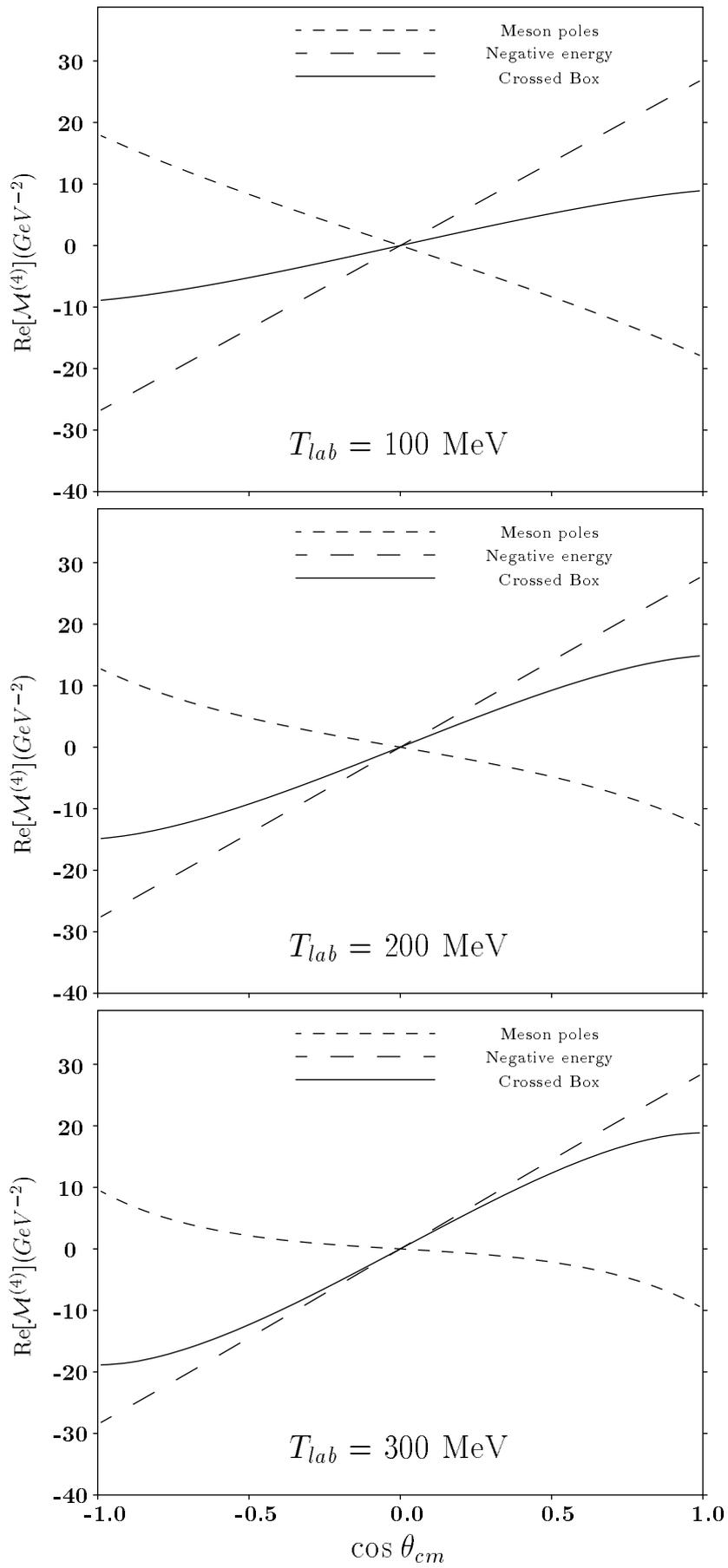
\end{document}